\documentclass[aps,prd,onecolumn,eqsecnum,amsmath,nofootinbib,preprintnumbers]{revtex4}%
\newcounter{example}[section]

\usepackage{appendix}
\usepackage{amsmath}
\usepackage{color,graphicx,float,subfigure}
\usepackage{amsfonts,amssymb,theorem,mathrsfs,times}
\usepackage{bm}
\usepackage{ulem}
\usepackage{amsfonts,amssymb,theorem,mathrsfs}
{\theorembodyfont{\upshape}
	}
{\theorembodyfont{\upshape}
	}
{\theorembodyfont{\upshape}
	}
{\theorembodyfont{\upshape}
	\newtheorem{dn}{Definition}}
{\theorembodyfont{\upshape}
	}
{\theorembodyfont{\upshape}
	}

\newcommand{\dalm}{\kern1pt\vbox{\hrule height 0.9pt\hbox{\vrule width
			0.9pt\hskip 2.5pt\vbox{\vskip 5.5pt}\hskip 3pt\vrule width
			0.3pt}\hrule height 0.3pt}\kern1pt}

\begin{document}
	\title{The particle surface of spinning test particles}
	
	%
	
	\author{ Yong Song\footnote{e-mail
			address: syong@cdut.edu.cn}}
	\author{Yiting Cen\footnote{e-mail
			address: 2199882193@qq.com}}
	\author{Leilei Tang\footnote{e-mail
			address: hamidun2233@foxmail.com}}
	\author{Jiabao Hu\footnote{e-mail
				address: 2463898559@qq.com}}
	\author{Kai Diao\footnote{e-mail
			address: kai\_diao@qq.com}}
	\author{Xiaofeng Zhao\footnote{e-mail
			address: 59403804@qq.com}}
	\author{Shunping Shi\footnote{e-mail
			address: shishunping13@cdut.edu.cn (corresponding author)}}

	
	\affiliation{
	College of Mathematics and Physics\\
	Chengdu University of Technology, Chengdu, Sichuan 610059,
	China}
	

	\date{\today}
	
	\begin{abstract}
	In this work, inspired by the definition of the photon surface given by Claudel, Virbhadra, and
	Ellis, we give an alternative quasi-local definition to study the circular orbits of single-pole particles. This definition does not only apply to photons but also to massive point particles. For the case of photons in spherically symmetric spacetime, it will give a photon surface equivalent to the result of Claudel, Virbhadra, and Ellis. Meanwhile, in general static and stationary spacetime, this
	definition can be regarded as a quasi-local form of the effective potential method. However, unlike
	the effective potential method which can not define the effective potential in dynamical spacetime,
	this definition can be applied to dynamical spacetime. Further, we generalize this definition
	directly to the case of pole-dipole particles. In static spherical symmetry spacetime, we verify the
	correctness of this generalization by comparing the results obtained by the effective potential
	method.
	\end{abstract}
	

	\maketitle


\section{Introduction}
Black hole is one of most important prediction of general relativity. To confirm its existence, scientists have made a lot of efforts. In 2019, the Event Horizon Telescope (EHT) Collaborations published the first images of a supermassive black hole at the center of the M87 galaxy~\cite{EventHorizonTelescope:2019dse}. Later, in 2021, the EHT Collaborations released the polarized images of the black hole~\cite{EventHorizonTelescope:2021bee,EventHorizonTelescope:2021srq}. Very recently, the EHT announced the image of the Galactic Center
Supermassive Black Hole~\cite{EventHorizonTelescope:2022xnr}.

To analysis these images, it is important to study the geodesic circular orbits of the astrophysical black holes~\cite{EventHorizonTelescope:2019pgp,EventHorizonTelescope:2022xqj}. On the one hand, to study the black hole shadow, we first need to study the circular orbits of photons. In static spherically symmetric spacetime, the photon sphere, where the location of the circular photon orbits, describes the boundary of the black hole shadow which corresponding to the shaded part of the image~\cite{Synge:1966okc,Virbhadra:1999nm}. On the other hand, in black hole accretion disk theory, the circular geodesic motion in the equatorial plane is of fundamental importance, the detail one can see~\cite{Abramowicz:2011xu}. The luminous part of the image corresponds to the accretion disk which is located at the stable circular orbit of the black hole~\cite{Abramowicz:2011xu,Cardoso:2008bp}.

In static spacetime and stationary spacetime, one can solve the geodesic equations and define the effective potential of the system to get the circular orbits of a timelike or null geodesic. However, astrophysical black holes in reality involve evolution, and one can not use this method to get the circular orbits because the effective potential of a dynamical system can not be defined. So, in recent years, the quasi-local studies of photon sphere, photon surface and their generalization definitions have attracted some attention. The first quasi-local definition of the photon surface is given by Claudel, Virbhadra, and Ellis which based on the umbilical hypersurface~\cite{Claudel:2000yi}. Based on this definition, they studied the photon surfaces in general spherically symmetric spacetimes. However, there are some problems in this definition~\cite{Cao:2019vlu}: (i). The definition allows that spacetime, which in the absence of gravity, exists a photon surface. (ii). From their definition, one can not to get the boundary condition when solving the equation of the photon surface. (iii). The umbilical condition, i.e., the shear tensor of a hypersurface is vanishing, is too restrictive and makes their definition does not work in an axisymmetric stationary spacetime. (iv). Their definition is for photon, so, it can not deal with the case of massive point particle. The problem (i) and (ii) has been solved by~\cite{Cao:2019vlu} in general spherically symmetric spacetimes which based on the codimension-2 surface of the spacetime. For the problem (iii), there are many generalized studies. Such as Yoshino et al. generalize the photon surface to be a loosely trapped surface~\cite{Shiromizu:2017ego} and (dynamically) transversely trapping surface~\cite{Yoshino:2017gqv,Yoshino:2019dty,Galtsov:2019bty}; Kobialko et al. generalize the photon surface to be a fundamental photon hypersurfaces and fundamental photon regions~\cite{Kobialko:2020vqf,Kobialko:2021aqg}. The problem (iv) has been solved by~\cite{Kobialko:2022uzj} which generalize the photon surface to massive particle surface and~\cite{Song:2022fdg} which generalize the photon surface to (partial) particle surface.

In this paper, we ask and discuss some questions about the quasi-local study of circular orbits. Firstly, in static and stationary spacetimes, one always uses the effective potential method to obtain the circular orbits of the spacetime. However, the method given by Claudel, Virbhadra, and Ellis for quasi-local study of circular orbits seems to have no connection with the effective potential method. So, is there any relationship between them? Secondly, previous work all consider the point particles, but in reality, particles may have intrinsic properties, such as spin. The circular orbits of a spinning test particle in different spacetimes have been studied~\cite{Jefremov:2015gza,Toshmatov:2019bda,Toshmatov:2020wky}. But, how to study the circular orbit of spinning test particles quasi-locally is still a question worth studying. Here, we only focus on spinning extended test bodies up to pole-dipole order. Unlike the point particle or the single-pole particle, it does not satisfy the geodesic equation but the Mathisson-Papapetrous-Dixon (MPD) equation~\cite{Mathisson:1937zz,Papapetrou:1951pa,Tulczyjew:1959,Taub:1964,Madore:1969,Dixon:1964,Dixon:1965:1,Dixon:1970:1,Dixon:1970:2,Dixon:1973:1,Dixon:1974:1,Dixon:1979,Obukhov:2015eqa,Semerak:1999qc,Hackmann:2014tga,Steinhoff:2009tk}. A direct generalization of the quasi-local definition given by~\cite{Claudel:2000yi,Song:2022fdg} is hard to check its correctness, and we will discuss this point in Appendix A. In this paper, we give an alternative definition to study the circular orbit of a single-pole particle which is equivalent  to the definition given by~\cite{Claudel:2000yi,Song:2022fdg}. This definition can be regarded as a quasi-local form of the effective potential method in general static and stationary spacetime, and can be easily generalized to the case of pole-dipole particles. 

This paper is organized as follows: In section \ref{section2}, we will give the quasi-local definition of a single-pole particle surface and study its circular orbits in static spherical symmetric spacetime by using the effective method and the quasi-local definition. Further, as an example in dynamical spacetime, we will study the photon surface in vaidya spacetime by using this definition. In section \ref{section3}, at first, we will give a brief review of the equations of motion of spinning extended test bodies, i.e., MPD equations. Then, we will generalize the quasi-local definition of the single-pole particle surface to pole-dipole particle surface. At last, we will demonstrate the equivalence of the quasi-local definition and the effective potential method in static spherical symmetic spacetime. Section \ref{conclusion} is devoted to the conclusion and  discussion. In Appendix A, we will give a disussion about the definition of the pole-dipole particle surface based on the definition given by~\cite{Claudel:2000yi,Song:2022fdg}.

Convention of this paper: We choose the system of geometrized unit, i.e., set $G=c=1$. We use the symbol $(\mathcal{M},\nabla_a,g_{ab})$ to denote a manifold $\mathcal{M}$ with metric $g_{ab}$ and covariant derivative operator $\nabla_a$, and ($g_{ab},\nabla_a$) satisfies the compatibility condition, i.e., $\nabla_a g_{bc}=0$. The abstract index formalism has been used to clarify some formulas or calculations. The curvature $R_{abcd}$ of the spacetime is defined by $R_{abcd}v^d=(\nabla_a\nabla_b-\nabla_b\nabla_a)v_c\,$ for an arbitrary tangent vector field $v^a$.

\section{Single-pole particles}\label{section2}
\subsection{Quasi-local definition}
Inspired by the definition given by~\cite{Claudel:2000yi,Song:2022fdg}, we give the following quasi-local definition of the single-pole particle surface:

\dn{ \it {Let $(\mathcal{S},D_a,h_{ab})$ be a timelike hypersurface (or a subset of a timelike hypersurface) of $(\mathcal{M},\nabla_a,g_{ab})$. Let $v^a$ be a unit normal vector to $\mathcal{S}$, i.e., it satisfies $v^av_a=1$. The metric $g_{ab}$ can be decomposed as
\begin{align}
	g_{ab}=h_{ab}+v_av_b\;,
\end{align}
Let $\gamma$ be the geodesic of a single-pole particle that intersects $\mathcal{S}$ at point $p$. At point $p$, the tangent vector $K^a$ of the geodesic can be decomposed as
\begin{align}
	\label{kdecompose}
	K^a=K_{\parallel}^a+K^a_{\perp}=h^a{}_bK^b+v^av_bK^b=k^a+v^av_bK^b\;.
\end{align}
where $k^a\equiv K_{\parallel}^a=h^a{}_bK^b$ is parallel to $\mathcal{S}$ and $K^a_{\perp}=v^av_bK^b$ is normalized to $\mathcal{S}$. If for $\forall p\in S$, there exists at least one $\gamma\in S$ passing through $p$ and satisfies
\begin{align}
\label{def1}
K^av_a|_p=0\;,
\end{align}
and
\begin{eqnarray}
\label{def2}
K^b\nabla_b (k^ak_{a})|_p=0\;,
\end{eqnarray}
Then $\mathcal{S}$ is called a (partial) single-pole particle surface.
}}

\dn{ \it {A (partial) single-pole particle surface is called stable if it satisfies $v^c\nabla_c[K^b\nabla_b (k^ak_{a})]|_p\ge 0$, and unstabe if it satisfies $v^c\nabla_c[v^b\nabla_b(k^a k_{a})]|_p<0$.}}

\vspace{0.3cm}

Below, we give some remarks:

(i). Here, we have assumed that $v^a$ is out pointing. Roughly speaking, the out pointing requirement refers to a direction from the center of the system to infinity.

(ii). From the geodesic equation $K^a\nabla_aK^b=0$, one can always have $K^b\nabla_b(K^aK_a)=0$. But, $K^b\nabla_b(k^a k_{a})$ is not vanished in general, and condition (\ref{def2}) will give a non-trivial condition.

(iii). This definition is equivalent to the definition of the particle surface given by~\cite{Song:2022fdg}. The condition of particle surface in~\cite{Song:2022fdg} is $k^ak^bK_{ab}=0$\footnote{In the definition given by~\cite{Song:2022fdg}, $k^a$ or $K^a$ does not matter, because the projection operator $h^a{}_b$ is implied in $K_{ab}$. But, in our definition, one must distiguish between $K^a$ and $k^a$.}, where $K_{ab}=h_a{}^ch_b{}^d\nabla_cv_d$ is the second fundamental form of $\mathcal{S}$. Then, for $\forall p\in\mathcal{S}$, we have
\begin{align}
k^ak^bK_{ab}&=k^ak^b\nabla_av_b\nonumber\\
&=(K^a-v^av_cK^c)(K^b-v^bv_dK^d)\nabla_av_b\nonumber\\
&=K^aK^b\nabla_av_b=K^a\nabla_a(v_bK^b)\nonumber\\
&=K^c\nabla_c(\sqrt{K^aK_a-k_ak^a})\nonumber\\
&=\frac{1}{2\sqrt{K^aK_a-k_ak^a}}K^c\nabla_c(K^aK_a-k_ak^a)\nonumber\\
&=-\frac{1}{2v^aK_a}K^b\nabla_b(k^ak_a)\;,
\end{align}
where we have used eq.(\ref{kdecompose}), (\ref{def1}) and the geodesic equation $K^a\nabla_aK^b=0$. So, when $k^ak^bK_{ab}=0$, we have $K^b\nabla_b(k^ak_a)=0$.

(iv). In static and stationary spacetime, this definition will be equivalent to effective potential method, and we will illustrate the equivalence in static spherical symmetric spacetime in the next subsection. For the case of stationary spacetime, a similar argument follows.

(v). Definition 2 is a direct modification of the stability condition given by~\cite{Song:2022fdg}. From the condition $v^c\nabla_c[K^b\nabla_b (k^ak_{a})]=0$, one can get the innermost stable circular orbits.

(vi). The above definitions may have a widely application and they can be easily generalized to other situations. In next section, we will generalize this definition to the case of pole-dipole particles, and study the circular orbits of a pole-dipole particle in static spherical symmetric spacetime.

\subsection{The circular orbits of single-pole particle in static spherical symmetric spacetime}
The metric of the general static spherical symmetric spacetimes in the $\{t,r,\theta,\phi\}$ coordinates can be written as
\begin{eqnarray}
	\label{metricstatic}
	ds^2=-F(r)dt^2+H(r)dr^2+r^2(d\theta^2+\sin^2\theta d\phi^2)
\end{eqnarray}
where $F$ and $H$ are functions of radial coordinate $r$. Considering the untraped region, we have $F(r)>0$ and $H(r)>0$. Due to the spherical symmetry of the metric (\ref{metricstatic}), we can focus our analysis on the equatorial plane, i.e., $\theta=\pi/2$. A single-pole particle that behaves like a point particle follows a geodesic trajectory in spacetime. The four-velocity of the geodesic can be expressed as
\begin{eqnarray}
	K^a=\frac{dx^\mu(\lambda)}{d\lambda}\bigg(\frac{\partial}{\partial x^\mu}\bigg)^a=\frac{dt(\lambda)}{d\lambda}\bigg(\frac{\partial}{\partial t}\bigg)^a+\frac{dr(\lambda)}{d\lambda}\bigg(\frac{\partial}{\partial r}\bigg)^a+\frac{d\phi(\lambda)}{d\lambda}\bigg(\frac{\partial}{\partial \phi}\bigg)^a\;,
\end{eqnarray}
where $\lambda$ is the parameter of the geodesic. The normalized condition of the four-velocity is
\begin{eqnarray}
K^aK_a=\delta\;,
\end{eqnarray}
where $\delta=0$ for a photon and $\delta=-1$ for a massive point particle. Then, we have
\begin{eqnarray}
	\label{KaKa}
	-F(r)\bigg(\frac{dt}{d\lambda}\bigg)^2+H(r)\bigg(\frac{dr}{d\lambda}\bigg)^2+r^2\bigg(\frac{d\phi}{d\lambda}\bigg)^2=\delta.
\end{eqnarray}
Along the geodesic, there are two conserved quatities, i.e.,
\begin{eqnarray}
	\label{eeff}
	&&e=K_a\bigg(\frac{\partial}{\partial t}\bigg)^a=-F\bigg(\frac{dt}{d\lambda}\bigg)\;,\\
	\label{leff}
	&&l=K_a\bigg(\frac{\partial}{\partial \phi}\bigg)^a=r^2\bigg(\frac{d\phi}{d\lambda}\bigg)\;.
\end{eqnarray}
where $e$ is the conserved orbital energy and $l$ is the conserved orbital angular momentum of the geodesic. 
\subsubsection{Effective potential method}
Considering eq.(\ref{eeff}) and (\ref{leff}), eq.(\ref{KaKa}) can be reformulated as
\begin{eqnarray}
\label{dr}
\bigg(\frac{dr}{d\lambda}\bigg)^2=\frac{1}{H}\bigg(\frac{e^2}{F}-\frac{l^2}{r^2}+\delta\bigg)\;,
\end{eqnarray}
Then, the effictive potential can be defined as~\cite{Cardoso:2008bp}
\begin{eqnarray}
\label{Veff}
V_{\mathrm{eff}}=\frac{1}{H}\bigg(\frac{e^2}{F}-\frac{l^2}{r^2}+\delta\bigg)\;.
\end{eqnarray}
The particle moves along a circular orbit when two conditions are satisfied simultaneously which has been pointed out in~\cite{Jefremov:2015gza,Toshmatov:2020wky}:
\begin{itemize}
\item[(1).]  The particle has zero radial velocity, i.e.
\begin{eqnarray}
\label{eff1}
\frac{dr}{d\lambda}=0\;.
\end{eqnarray}
\item [(2)] The particle has zero radial acceleration, i.e.
\begin{eqnarray}
	\label{eff2}
	\frac{d^2r}{d\lambda^2}=0,\quad\Longrightarrow \frac{dV_{\mathrm{eff}}}{dr}=0\;.
\end{eqnarray}
\end{itemize}
It should be noted that condition (\ref{eff1}) and (\ref{eff2}) only need to hold at one point. As long as the particle has zero radial velocity and radial acceleration at one point,  its trajectory will be a circular orbit.

The stability condition of a circular orbit is
\begin{equation}
	\label{stabilityeff}
	\frac{d^2V_{\mathrm{eff}}}{dr^2}=\begin{cases}
		<0\quad\quad & \textrm{unstable}\;,\\
		\\
		\ge 0\quad\quad & \textrm{stable}\;.
	\end{cases}
\end{equation}
When $d^2V_{\mathrm{eff}}/dr^2=0$, the circular orbits correspond to the innermost stable circular orbit (ISCO). 

From eq.(\ref{eff1}), we can get
\begin{eqnarray}
\label{11}
e^2-\frac{Fl^2}{r^2}+F\delta=0\;.
\end{eqnarray}
And frome eq.(\ref{eff2}), we have
\begin{eqnarray}
\label{21}
\frac{2Fl^2}{r^3}+\bigg(\delta-\frac{l^2}{r^2}\bigg)F'=0\;.
\end{eqnarray}
Combining eq.(\ref{11}) and (\ref{21}), we can get the equation of the circular orbits in the general static spherical symmetric spacetimes as follows
\begin{eqnarray}
\label{effcircularorbit}
e^2=\frac{2\delta F^2}{rF'-2F}\;,\quad l^2=\frac{\delta r^3 F'}{rF'-2F}\;.
\end{eqnarray}
where a prime denotes a derivative with respect to areal radius $r$. The stability condition becomes
\begin{displaymath}
	\sqrt{\frac{\delta F^2}{-4F+2rF'}}\frac{-2rF'^2+3FF'+rFF''}{rF^2}= \left\{ \begin{array}{ll}
		<0\quad\quad & \textrm{unstable}\;,\\
		\\
		\ge 0\quad\quad & \textrm{stable}\;.
	\end{array} \right.
\end{displaymath}

For a photon, the circular orbits can not be directly derived from the above equations\footnote{For the case of null geodesic, the eq.(\ref{effcircularorbit}) and the stability condition will become the form of $0/0$. Here, we just write it in a uniform form so that we can compare it with the subsequent results easily.}, it should be derived by solving eq.(\ref{11}) and (\ref{21}) which have been set $\delta=0$. Then, one can get the circular orbit of a photon satisfies
\begin{eqnarray}
2F=rF^\prime,\quad \frac{e^2}{l^2}=\frac{F}{r^2}\;.
\end{eqnarray}
And the stability condition of a photon circular orbits can be reduced to
\begin{displaymath}
	r^2F''-2F= \left\{ \begin{array}{ll}
		<0\quad\quad & \textrm{unstable}\;,\\
		\\
		\ge 0\quad\quad & \textrm{stable}\;.
	\end{array} \right.
\end{displaymath}
For a massive point particle, the circular orbits satisfies
\begin{eqnarray}
e^2=\frac{2F^2}{2F-rF'},\quad l^2=\frac{r^3F'}{2F-rF'}\;.
\end{eqnarray}
And the stability condition in this case can be reduced to
\begin{displaymath}
	3FF'/r-2(F')^2+FF''= \left\{ \begin{array}{ll}
		<0\quad\quad & \textrm{unstable}\;,\\
		\\
		\ge 0\quad\quad & \textrm{stable}\;.
	\end{array} \right.
\end{displaymath}
\subsubsection{Quasi-local method}
In static spherical symmetric spacetime, there exists a family of circular orbits with specific parameters located at a hypersurface which has $r=\mathrm{constant}$. The normal vector $v^a$ of this hypersurface can be written as
\begin{eqnarray}
v^a=\sqrt{H}\bigg(\frac{\partial}{\partial r}\bigg)^a\;,
\end{eqnarray}
Consider a timelike or null geodesic intersects the hypersurface at point $p$ and its tangent vector $K^a$ in equatorial plane can be written as
\begin{eqnarray}
K^a=\{k^t,k^r,0,k^\phi\}=\{-\frac{e}{F},\frac{dr}{d\lambda},0,\frac{l}{r^2}\}
\end{eqnarray}
where we have used eq.(\ref{eeff}) and (\ref{leff}). From the condition(\ref{def1}), at point $p$, we have
\begin{eqnarray}
\label{Kavastatic}
K^av_a|_p=\frac{dr}{d\lambda}\sqrt{H}=0\;,
\end{eqnarray}
Then we get the equation $k^r=dr/d\lambda=0$ which corresponding to the the first condition (\ref{eff1}) of the circular orbits. The vector $k^a$ can be written as
\begin{eqnarray}
	k^a=\{-\frac{e}{F},0,0,\frac{l}{r^2}\}\;.
\end{eqnarray}
And from the condition(\ref{def2}), at point $p$, we have
\begin{align}
K^b\nabla_b(k^a k_{a})|_p=k^r\frac{\partial}{\partial r}\bigg(-\frac{e^2}{F}+\frac{l^2}{r^2}\bigg)=Hk^r\frac{\partial}{\partial r}\bigg[\frac{1}{H}\bigg(\frac{e^2}{F}-\frac{l^2}{r^2}+\delta\bigg)\bigg]=Hk^r\frac{dV_{\mathrm{eff}}}{dr}\;,
\end{align}
where we have used the eq.(\ref{Veff}) and eq.(\ref{Kavastatic}). Thus, we get
\begin{align}
\frac{dV_{\mathrm{eff}}}{dr}=0\;,
\end{align}
which corresponds to the second condition (\ref{eff2}) of the circular orbits. Further, from definition 2, one can easily get the stability condition of the circular orbits which is the same as (\ref{stabilityeff}).

\subsection{Dynamic spacetime}
As an example in dynamical spacetime, we study the evolution of the null circular orbits in Vaidya spacetime. For the case of a timelike geodesic, the calculation can be performed similarly. 

In the in-going null coordinate $\{v,r,\theta,\phi\}$, the metric of the 4-dimensional Vaidya spacetime can be written as~\cite{Vaidya:1951zza}
\begin{eqnarray}
	ds^2=-\bigg(1-\frac{2M(v)}{r}\bigg)dv^2+2dvdr+r^2(d\theta^2+\sin^2\theta d\phi)\;,
\end{eqnarray}
where $M(v)$ is a freely specifiable function of $v$. The unit normal vector $v^a$ of the particle surface in Vaidya spacetime can be written as~\cite{Claudel:2000yi,Song:2022fdg}
\begin{eqnarray}
v^a=\frac{1}{\sqrt{1-2M(v)/r-2\dot{r}}}\bigg(\frac{\partial}{\partial v}\bigg)^a+\frac{1-2M(v)/r-\dot{r}}{\sqrt{1-2M(v)/r-2\dot{r}}}\bigg(\frac{\partial}{\partial r}\bigg)^a\;.
\end{eqnarray}
where $``\cdot"$ stands for the derivative with respect to the coordinate time $``v"$. Due of the spherical symmetry of the system, we can also focus the analysis on the equatorial plane, i.e., $\theta=\pi/2$. The component of the tangent vector of a geodesic can be supposed as $K^a=\{k^t,k^r,0,k^\phi\}$, where $k^t\;,k^r\;,k^\phi\;$  can be considered as functions of $v$. Along the geodesic, the orbital angular momentum $l$ is conserved, so, we have
\begin{eqnarray}
	\label{lvaidya}
	&&l=K_a\bigg(\frac{\partial}{\partial \phi}\bigg)^a=r^2k^\phi\;.
\end{eqnarray}
From eq.(\ref{def1}), we have
\begin{eqnarray}
\label{vaidyakv}
K^av_a|_p=\dot{r}k^t-k^r=0\;.
\end{eqnarray}
At point $p$, from the normalized condition of $K^a$, we have
\begin{align}
\label{vaidyaKK}
K^aK_a|_p=-\bigg(1-\frac{2M(v)}{r}\bigg)(k^t)^2+2k^tk^r+\frac{l^2}{r^2}=0\;,
\end{align}
and the geodesic equation $K^a\nabla_aK^b=0$ will give
\begin{align}
\label{vaidyage1}
-\frac{l^2}{r^3}+k^t\bigg[\frac{M(v)}{r^2}k^t+\dot{k}^t\bigg]&=0\;,\\
\label{vaidyage2}
-[l^2+2(k^t)^2M(v)^2]r+M(v)[2l^2+r^2(1-2\dot{r})(k^t)^2]+r^3k^t[r\dot{k}^t+M(v)k^t]&=0\;,
\end{align}
The vector $k^a$ can be expressed as $\{k^t,\dot{r}k^t,0,l/r^2\}$. Combining eq.(\ref{vaidyakv}), (\ref{vaidyaKK}), (\ref{vaidyage1}) and (\ref{vaidyage2}), we get
\begin{align}
\label{ktkr}
&k^r=\dot{r}k^t\;,\\
\label{kt}
&k^t=\frac{1}{\sqrt{r[r-2r\dot{r}-2M(v)]}}\;,\\
\label{kr}
&\dot{k}^t=\frac{[r-3M(v)-2r\dot{r}]l}{r^2\sqrt{r[r-2r\dot{r}-2M(v)]}}\;.
\end{align}
From eq.(\ref{def2}), we have
\begin{align}
\label{vaidyaps}
\frac{-2[l^2+rM(v)(k^t)^2]k^r-2r^2[r-2M(v)-2r\dot{r}](k^t)^2\dot{k}^t+[\dot{M}(v)+r\ddot{r}]k^t}{r^3}=0\;.
\end{align}
Putting eq.(\ref{ktkr}), (\ref{kt}) and (\ref{kr}) into eq.(\ref{vaidyaps}), we finally get
\begin{align}
\ddot{r}=\frac{1}{r}\bigg[\bigg(1-\frac{3M(v)}{r}\bigg)\bigg(1-\frac{2M(v)}{r}-3\dot{r}\bigg)-\dot{M}(v)+2\dot{r}^2\bigg]
\end{align}
Which is exactly equation of the photon surface in~\cite{Claudel:2000yi,Song:2022fdg}.
\section{Pole-dipole particle}\label{section3}
\subsection{Review the Mathisson-Papapetrous-Dixon (MPD) equations}
In this section, we will give a brief review of the Mathisson-Papapetrous-Dixon  (MPD) equations. For more details, one can find in the references~\cite{Mathisson:1937zz,Papapetrou:1951pa,Dixon:1964,Dixon:1965:1,Dixon:1970:1,Dixon:1970:2,Dixon:1973:1,Dixon:1974:1,Dixon:1979}.

The equations of motion of spinning extended test bodies up to the pole-dipole order are given by the MPD equations which read
\begin{eqnarray}
	\label{MPD1}
	&&\dot{P}^a=-\frac{1}{2}R^a{}_{bcd}u^b S^{cd}\;,\\
	\label{MPD2}
	&&\dot{S}^{ab}=2P^{[a}u^{b]}\;.
\end{eqnarray}
where $u^a=dx^a/ds$ is the 4-velocity of the body along its world line, and the dot denotes the covariant derivative with respect to the proper time $``s"$, i.e., $``\cdot"=D/ds=u^a\nabla_{a}$. The antisymmetric tensor $S^{ab}$ is the spin tensor and $P^a$ is 4-momentum of the test body. 

In order to close the system of eqs. (\ref{MPD1}) and (\ref{MPD2}), a supplementary condition has to be imposed. In this work, to restrict the spin tensor to generate rotations only, we focus on the Tulczyjew spin-supplementary condition~\cite{Tulczyjew:1959}, i.e.,
\begin{eqnarray}
	\label{TC}
	S^{ab} P_{b}=0\;.
\end{eqnarray}
From eq.(\ref{TC}), it turns out that the canonical momentum and the spin of the body provide two independent conserved quantities given by the relations~\cite{Semerak:1999qc,Hackmann:2014tga}
\begin{eqnarray}
	\label{ppm2}
	&&P^a P_{a}=-M^2\;,\\
	\label{SSs2}
	&&\frac{1}{2}S^{ab}S_{ab}=S^2\;,
\end{eqnarray}
where $M$ is the ‘dynamical’, ‘total’ or ‘effective’ rest mass of the body and $S$ is the spin length of the body.
The spin four-vector can be defined as
\begin{eqnarray}
	S^a=\frac{1}{2M}\epsilon^{ba}{}_{cd}P_b S^{cd}\;,
\end{eqnarray} 
where $\epsilon^{ba}{}_{cd}$ is the Levi–Civita tensor. It is easy to find out $S^a$ is orthogonal to $P^a$, i.e., $S^aP_a=0$. In addition to the conserved quantities resulting from
the Tulczyjew condition, there exist also the conserved quantities associated to the spacetime symmetries given by the Killing vectors $\xi^\mu$, which can be expressed as
\begin{eqnarray}
	\label{conserved}
	P^a\xi_a-\frac{1}{2}S^{ab}\nabla_b\xi_{a}=P^a\xi_a-\frac{1}{2}S^{ab}\partial_b\xi_{a}=\mathrm{constant}\;.
\end{eqnarray}

\subsection{Quasi-local definition}
In this section, we will generalize the quasi-local definition of the single-pole particle surface to pole-dipole particle surface. The surface where the circular orbits of a pole-dipole particle located can be defined as:
\dn{ \it {Let $(\mathcal{S},D_a,h_{ab})$ be a timelike hypersurface (or a subset of a timelike hypersurface) of $(\mathcal{M},\nabla_a,g_{ab})$. Let $v^a$ be a unit normal vector to $\mathcal{S}$. The metric $g_{ab}$ can be decomposed as
\begin{align}
g_{ab}=h_{ab}+v_av_b\;,
\end{align}
Let $x^a(s)$ is the world line of a pole-dipole particle that intersects $\mathcal{S}$ at point $p$. Let $P^a$ be the 4-momentum of the pole-dipole particle and can be decomposed as
\begin{align}
\label{Pdecompose}
P^a=P_{\parallel}^a+P^a_{\perp}=h^a{}_bP^b+v^av_bP^b=p^a+v^av_bP^b\;.
\end{align}
where $p^a\equiv P_{\parallel}^a=h^a{}_bP^b$ is parallel to $\mathcal{S}$ and $P^a_{\perp}=v^av_bP^b$ is normalized to $\mathcal{S}$. If for $\forall p\in S$, there exists at least one $x^a(s)$ passing through $p$ and satisfies
\begin{align}
\label{def3}
P^av_a|_p=0\;,
\end{align}
and
\begin{eqnarray}
\label{def4}
P^b\nabla_b (p^ap_{a})|_p=0\;,
\end{eqnarray}
Then $\mathcal{S}$ is called a pole-dipoe particle surface.
}}

\dn{ \it {A pole-dipoe particle surface is called stable if it satisfies $v^c\nabla_c[P^b\nabla_b (p^ap_{a})]|_p\ge 0$, and unstabe if it satisfies $v^c\nabla_c[P^b\nabla_b (p^ap_{a})]|_p<0$.}}

Below, we give some remarks:

(i). This definition is a direct generalization of the definition of single-pole particle surface. In this work, for simplicity, we only focus on the static spherical symmetric spacetime.

(ii). Definition 3 is based on the Tulczyjew spin-supplementary condition. In static spherical symmetric spacetime, spin can be chosen to be orthogonal to the equatorial plane, i.e., the spin four-vector $S^a$ perpendicular to equatorial plane~\cite{Mohseni:2010rm,Hojman:2018evi}. This make sure that the motion of the pole-dipole particle is planar~\cite{Rietdijk:1992tx}.



\subsection{The equivalence between the effective potential method and the quasi-local definition}\label{section5}
In this section, we will illustrate the equivalence between the effective potential method and the quasi-local definition to study the circular orbits of a pole-dipole particle in the general static spherical symmetric spacetimes.

Because of the spherical symmetry of the line element (\ref{metricstatic}), we can choose the equatorial plane, i.e., $\theta=\pi/2$. And we can suppose 4-momentum of the pole-dipole particle to be $P^a=\{p^t,p^r,0,p^\phi\}$, where $p^t\;,p^r\;,p^\phi\;$ can be considered as functions of $r$. Along the world line, there are two conserved quantities for the pole-dipole particle, i.e., the energy $E$ and the angular momentum $L$. From eq.(\ref{conserved}), the conserved quantities can be expressed as 
\begin{eqnarray}
	\label{E}
	&&-E=p_t+\frac{1}{2}F^\prime S^{tr}\;,\\
	\label{L}
	&&L=p_\phi+r S^{r\phi}\;.
\end{eqnarray}
where a prime denotes the derivative with respect to radial coordinate $r$. From the Tulczyjew spin supplementary condition (\ref{TC}), we have
\begin{eqnarray}
\label{Stphi}
&&S^{t\phi}=-\frac{p_r}{p_\phi}S^{tr}\;,\\
\label{Srphi}
&&S^{r\phi}=\frac{p_t}{p_\phi}S^{tr}\;.
\end{eqnarray}
 From eq.(\ref{ppm2}), we can get
\begin{eqnarray}
	\label{pr}
	(p^r)^2=\frac{1}{FH}p_t^2-\frac{1}{H}\bigg(\frac{p_\phi^2}{r^2}+\mathcal{M}^2\bigg)\;.
\end{eqnarray}
Combining eq.(\ref{Stphi}), (\ref{Srphi}) and (\ref{pr}), from the spin conservation eq.(\ref{SSs2}), we have
\begin{eqnarray}
S^{tr}=\frac{sp_\phi}{\sqrt{FH}r}\;,
\end{eqnarray}
where $s=S/\mathcal{M}$ is specific spin parameter. It should be noted
that $s$ can have both negative and positive values depending on the direction of spin with respect to direction of $p_\phi$. From the conservation of energy (\ref{E}) and angular momentum (\ref{L}), we have
\begin{eqnarray}
\label{pt}
&&p_t=-\frac{2rFHE+F^\prime\sqrt{FH}sL}{2rFH-F^\prime s^2}\;,\\
\label{pphi}
&&p_\phi=\frac{2r[FHL+\sqrt{FH}sE]}{2rFH-F^\prime s^2}\;.
\end{eqnarray}
\subsubsection{Effective potential method}
In this subsection, we will review the effective potential method to get the circular orbits of the pole-dipole particles~\cite{Toshmatov:2020wky}. 

Putting eq.(\ref{pt}) and (\ref{pphi}) into eq.(\ref{pr}), we get the result that
\begin{eqnarray}
\label{pr2}
(p^r)^2=A(E-V_+)(E-V_-)\;,
\end{eqnarray}
where
\begin{eqnarray}
A=\frac{4F(r^2H-s^2)}{(2FHr-F^\prime s^2)^2}
\end{eqnarray}
and
\begin{eqnarray}
\label{Vpm}
V_{\pm}=\frac{(2F-rF^\prime)\sqrt{FH}sL}{2(FHr^2-Fs^2)}\pm\frac{F^\prime s^2-2FHr}{FH(s^2-Hr^2)}\sqrt{H(L^2+M^2r^2)-M^2s^2}\;.
\end{eqnarray}
which is consistent with the result in~\cite{Toshmatov:2020wky}. According to eq.(\ref{pr2}), the energy of the particle must satisfy the conditions
\begin{eqnarray}
E\ge V_+\;,\quad \mathrm{or}\quad E\le V_-\;,
\end{eqnarray}
in order to have $(p^r)^2\ge 0$. Below, we focus on the case of the pole-dipole particle with positive energy which coincides with the effective potential to be $V_{\mathrm{eff}}=V_+$. Combining eq.(\ref{eff1}) and (\ref{eff2}), one can get the circular orbits of a pole-dipole particle on the equatorial plane in static spherical symmetric spacetime.

\subsubsection{Quasi-local study of the circular orbits}
In this subsection, we will use the definition 2 to study the circular orbits of a pole-dipole particle in static spherical symmetric spacetime.

In static spherical symmetric spacetime, the circular orbits of the pole-dipole particles are not evolved, and they satisfy the condition that $r_o=\mathrm{constant}$, where $r_o$ is the location of the circular orbit. Then, the normal vector $v^a$ of the particle surface can be written as
\begin{eqnarray}
	v^a=\frac{1}{\sqrt{H}}\bigg(\frac{\partial}{\partial r}\bigg)^a\;.
\end{eqnarray}
From condition (\ref{def3}), we have
\begin{eqnarray}
\label{Pava}
	P^av_a|_p=p^r\sqrt{H}=0\;,
\end{eqnarray}
Then we get $p^r=0$. From eq.(\ref{ppm2}) and considering (\ref{Pava}), we have the following result
\begin{eqnarray}
\label{papa}
\frac{(4FHs^2-4FH^2r^2)E^2+(8FH\sqrt{FH}sL-4H\sqrt{FH}F^\prime rsL)E+4F^2H^2L^2-HF^{\prime 2}s^2L^2}{(2FHr-s^2F^\prime)^2}=-\mathcal{M}^2\;,
\end{eqnarray}
Organizing the above result, we get
\begin{eqnarray}
\label{EV}
B(E-V_+)(E-V_-)=0\;,
\end{eqnarray}
where $V_{\pm}$ have been given by eq.(\ref{Vpm}), and
\begin{eqnarray}
B=\frac{4FH(s^2-r^2H)}{(2FHr-s^2F^\prime)^2}\;.
\end{eqnarray}
And $p^a=\{p^t,0,0,p^\phi\}$. From eq.(\ref{def4}), we have
\begin{eqnarray}
P^b\nabla_b(p^ap_a)=P^b\partial_b(p_ap^a)=p^r\partial_r(p_\mu p^\mu)\;.
\end{eqnarray}
Considering the result in eq.(\ref{papa}) and (\ref{EV}), we can get
\begin{eqnarray}
\partial_r(p_\mu p^\mu)=\partial_r(p_\mu p^\mu+\mathcal{M}^2)=\partial_r[B(E-V_+)(E-V_-)]=-B(E-V_-)\frac{\partial V_+}{\partial r}=0\;,
\end{eqnarray}
Then, we have
\begin{eqnarray}
\frac{\partial V_+}{\partial r}=\frac{\partial V_{\mathrm{eff}}}{\partial r}=0\;.
\end{eqnarray}

\vspace{0.3cm}
Here, we make a summary of this section: In this section, we give definition 2, which is a direct generalization of definition 1, to study the circular orbits of a pole-dipole particle, and illustrated its equivalence with the effective potential method in the general static spherical symmetric spacetimes.

(1). The condition $dr/dt=0$ is equivalent to 
\begin{eqnarray}
p^av_a=0\;.
\end{eqnarray}

(2). The condition $dV_{\mathrm{eff}}/dr=0$ is equivalent to 
\begin{eqnarray}
P^b\nabla_b(p_ap^a)=0\;.
\end{eqnarray}
Further, it is not hard to find that the stability condition, i.e., $v^c\nabla_c[P^b\nabla_b(p_ap^a)]\ge 0$, is equivalent to the following condition
\begin{eqnarray}
	\frac{\partial^2 V_+}{\partial r^2}=\frac{\partial^2 V_{\mathrm{eff}}}{\partial r^2}\ge 0\;.
\end{eqnarray}
Using the condition that $v^c\nabla_c[P^b\nabla_b(p_ap^a)]=0$, one can get the ISCO of the pole-dipole particle in static spherical symmetric spacetime.

\section{Discussion and conclusion}\label{conclusion}
In the Appendix A, we give a possible definition of the pole-dipole particle surface based on the definition given by~\cite{Claudel:2000yi,Song:2022fdg}. If this generalized definition is right, it can be applied to quite general spacetimes. But by this definition, because of the complicated calculation of the result, it is hard to check its correctness even in Schwarzschild spacetime. So, we need to find an another quasi-local definition of the pole-dipole particle surface.

In this paper, by deforming the condition of the particle surface in~\cite{Claudel:2000yi,Song:2022fdg}, we obtained an alternative form of the quasi-local definition. Definition 1 can be regarded as a quasi-local form of the effective potential method in static and stationary spacetime. In dynamical spacetime, we verified its correctness by taking the example of Vaidya spacetime. Further, we generalized definition 1 into the case of pole-dipole particles and illustrated its equivalence to the effective potential method in static spherical symmetric spacetime.

At present, the study of the circular orbits of a spinning test particle is focus on static and stationary spacetime. Although the calculation may be very complicated, our definitions provide a method for solving the evolution of the circular orbit of a spinning particle in a dynamical spacetime and lay the foundation for studying the evolution of the accretion disk of an astrophysical black hole.

When solving the equation of the particle surface, there will be little difference between definition 1 and the definition in~\cite{Claudel:2000yi,Song:2022fdg}. If one use definition 1 to get the evolution equations of the circular orbits in a dynamical spacetime, one need to consider the geodesic equation. This will make the solving process a little more difficult. In this work, we only considered some special cases, but for the more general situation, it can be solved similarly.


\section*{Acknowledgement}

This project is supported by the Open Research Fund of Computational Physics Key Laboratory of Sichuan Province, Yibin University (Grant No. YBXYJSWL-ZD-2020-005). This work is also supported by the Student’s Platform for Innovation and Entrepreneurship Training Program (No. S202110616084).


\appendix
\section{Quasi-local definition of the pole-diople particle surface based on the concept of the photon surface}\label{AppendixA}
Based on the quasi-local definition of the photon surface given by~\cite{Claudel:2000yi}, one may give the following possible definition for a pole-dipole particle surface:
\dn{ \it {Let $\mathcal{S}$ be a timelike hypersurface of $(\mathcal{M},\nabla_a,g_{ab})$. Let $x^a(s)$ is the world line of a pole-dipole particle on $\mathcal{S}$ and $u^a=dx^a/ds$ is the 4-velocity of the body along its world line. Let $v^a$ be a unit normal vector to $\mathcal{S}$ and $p^a$ be the 4-momentum of the pole-dipole particle. If for $\forall p\in S$, there exists at least one $x^a(s)\in S$ passing through $p$ and satisfies
\begin{eqnarray}
	\label{Def3}
	-\frac{1}{2}R^a{}_{bcd}u^b S^{cd}v_a+p^au^b\nabla_b v_a=0\;,
\end{eqnarray}
where $S^{ab}$ is the spin tensor, then $\mathcal{S}$ called is a pole-dipoe particle surface.

\vspace{0.3cm}

Below, we give some discussion about this definition:

(i). The condition (\ref{Def3}) can be obtained by the following consideration: The condition of the photon surface in~\cite{Claudel:2000yi} can be expressed as
\begin{eqnarray}
	\label{ps}
	K_{ab}k^ak^b=0\;,
\end{eqnarray}
where $K_{ab}$ is the second fundamental form of the photon surface and $k^a$ is the tangent vector of a null geodesic. Consider the normal vector of the particle surface is $v^a$, eq.(\ref{ps}) can be obtained by
\begin{eqnarray}
	k^a\nabla_a(k^bv_b)=0\Rightarrow K_{ab}k^ak^b=0\;,
\end{eqnarray}
where we have used the geodesic equation of $k^a$. Then, for a pole-dipole particle, following a similar consideration, one may get the following condition for the pole-dipole particle surface,
\begin{eqnarray}
	u^b\nabla_b(p^av_a)=0 \Rightarrow -\frac{1}{2}R^a{}_{bcd}u^b S^{cd}v_a+p^au^b\nabla_b v_a=0\;,
\end{eqnarray}
where we have used the MPD equations. Further, Combining the relation~\cite{Semerak:1999qc,Obukhov:2010kn}
\begin{eqnarray}
	u^a=\frac{m}{M^2}\bigg(p^a+\frac{2S_{ab}R_{bcde}p^cS^{de}}{4M^2+R_{abcd}S^{ab}S^{cd}}\bigg)\;,
\end{eqnarray}
where $m:=-p^au_a$ is a scalar parameter (the ‘kinematical’ or ‘monopole’ rest mass of a particle), one can get the equation of the pole-dipole particle surface. 

(ii). This definition is a natural generalization of the condition of a photon surface. If this definition is correct, it holds not only for static spherically symmetric spacetimes, but also for arbitrary spacetimes. However, one can check that, although it can be utilized to obtain the closed equation of the pole-dipole particle surface, it is difficult to check whether the result is right or not even in Schwarzschild spacetime. So, in this paper, we used a different approach to study the pole-dipole particle surface.

\end{document}